# Growth efficiency as a cellular objective in *Eschericia coli*

Tommi Aho[1*], Juha Kesseli[1], Olli Yli-Harja[1], Stuart A. Kauffman[1,2]

[1] Department of Signal Processing, Tampere University of Technology, Korkeakoulunkatu 1, 33720 Tampere, Finland

[2] Complex Systems Center, University of Vermont, U.S.A

Author email addresses: tommi.aho@tut.fi, juha.kesseli@tut.fi, olli.yli-harja@tut.fi, stuart.kauffman@uvm.edu

[*] Corresponding author. Contact details: tommi.aho@tut.fi, tel. +358 40 1981304, fax +358 3 3115 4989

## **Abstract**

The identification of cellular objectives is one of the central topics in the research of microbial metabolic networks. In particular, the information about a cellular objective is needed in flux balance analysis which is a commonly used constrained-based metabolic network analysis method for the prediction of cellular phenotypes. The cellular objective may vary depending on the organism and its growth conditions. It is probable that nutritionally scarce conditions are very common in the nature and, in order to survive in those conditions, cells exhibit various highly efficient nutrient processing systems like enzymes. In this study, we explore the efficiency of a metabolic network in transformation of substrates to new biomass, and we introduce a new objective function simulating growth efficiency.

We examined the properties of growth efficiency using a metabolic model for *Eschericia coli*. We found that the maximal growth efficiency is obtained at a finite nutrient uptake rate. The rate is substrate-dependent and it typically does not exceed 20 mmol/h/gDW. We further examined whether the maximal growth efficiency could serve as a cellular objective function in metabolic network analysis, and found that cellular growth in batch cultivation can be predicted reasonably well under this assumption. The fit to experimental data was found slightly better than with the commonly used objective function of maximal growth rate.

Based on our results, we suggest that the maximal growth efficiency can be considered as a plausible optimization criterion in metabolic modeling for *E. coli*. In the future, it would be interesting to study growth efficiency as a cellular objective also in other cellular systems and under different cultivation conditions.

# 1. Introduction

Flux balance analysis (FBA; [9, 21]) has been successfully applied to genome-scale models of microorganisms in order to characterize their metabolic capabilities [16]. FBA makes it possible to simulate different growth phenotypes attained under different environmental conditions and genetic modifications. The analysis can be performed without kinetic parameters in biochemical reactions equations, but using only stoichiometric and thermodynamic constraints. Typically, an objective function, representing the true cellular objective, needs also to be determined in flux balance analysis.

FBA is one of the constraint-based modeling methods that are based on the steady state assumption. The assumption states that the concentrations of metabolites not freely exchangeable with the environment are in a steady state. Given a metabolic network of $m$ metabolites and $n$ reactions, the network structure and the stoichiometric coefficients in reactions can be expressed in an $m \times n$ stoichiometric matrix $S$. Using $S$ and the steady state assumption, it is possible to form the following set of equations to comprehensively characterize all the feasible metabolic flux distributions:

56
$$\frac{dc}{dt} = Sv = 0$$
$$v_i^{lb} \leq v_i \leq v_i^{ub} \quad i = 1,...,n$$
(1)

57 In Equation (1), *c* is a vector of concentrations of non-exchangeable metabolites, and *v* is

58 a vector of reaction rates. The lower and upper bounds of reaction rates are defined by

59 $v_i^{lb}$ and $v_i^{ub}$, respectively. The bounds for reaction rates can be used to constrain specific

60 reactions to be irreversible, and to constrain the substrate uptake rate that usually is an

61 important parameter in constraint-based metabolic modeling. In FBA it is often assumed

62 that microorganisms aim to maximize their growth rate [8]. Therefore, a specific reaction

63 is implemented to describe the generation of new biomass. In FBA the maximum rate of

64 this reaction is determined using Linear Programming under the constraints of Equation

65 (1). Currently, the maximal growth rate has established usage as an objective function but

66 the rationale for cells always pursuing at maximal growth remains debatable [8].

67 Therefore, the research for other possible objective functions continues active. Other

68 suggested functions include the maximization of ATP yield [19, 17], the minimization of

69 the overall intracellular flux [4, 3], the maximization of ATP yield per flux unit [5], the

70 maximization of biomass yield per flux unit [18], the minimization of glucose

71 consumption [15], the minimization of reaction steps needed to produce biomass [13], the

72 maximization of ATP yield per reaction step [18], the minimization of redox potential

73 [12], the minimization of ATP producing fluxes [12], and the maximization of ATP

74 producing fluxes [11, 6, 12]. Similarly to these studies, FBA provides the methodological

75 framework also for our study of growth efficiency.

In the present work, we define the concept of growth efficiency and hypothesize that *Eschericia coli* uses it as the cellular objective. Maximal growth efficiency as the cellular objective would allow bacteria to utilize substrates efficiently to the production of new biomass while producing only little amount of waste, heat, or other side-products. In this work we explore the properties of growth efficiency using a genome-wide metabolic model for *Eschericia coli* [7] and study whether the growth efficiency could be considered as a plausible cellular objective in phenotypic simulations.

## 2. The definition and calculation of growth efficiency

We define the growth efficiency $\eta$ as the growth rate $v_{bm}$ (i.e. biomass production rate) divided by the substrate uptake rate $v_s$ ($\eta = v_{bm} / v_s$). Because $v_{bm}$ is largely determined by $v_s$, and in the following analysis we specifically focus on the effects of $v_s$ to $\eta$, we now define so-called growth efficiency function as $\eta = \mathrm{H}(v_s)$ and explore its properties. This simplification ignores specific other factors affecting $\eta$ via $v_{bm}$ but the sensitivity of $\eta$ to these factors will also be examined. The key assumption in our approach is that under specific conditions bacteria actively work to tune the substrate uptake rate such that the growth efficiency $\eta$ will be maximized. That is, the bacteria aim at substrate uptake rate $v_s^*$ that is optimal in the sense of $v_s^* = \underset{v_s}{\arg\max}\, \mathrm{H}(v_s)$.

In order to characterize the properties of growth efficiency and to study its use as a cellular objective function, we apply the constraints of Equation (1) and set maximal $\eta$ as the objective in Flux Balance Analysis. The problem is a linear-fractional problem where $\eta$, that is the ratio of biomass production rate to the substrate uptake rate, is maximized:

$$\begin{aligned} \max \quad & \eta \\ s.t. \quad & Sv = 0 \\ & v_{irr} \geq 0 \\ & v_i^{lb} \leq v_i \leq v_i^{ub} \quad \forall i \end{aligned} \quad (2)$$

We used the metabolic model iAF1260 for *Eschericia* coli [2] to study growth efficiency. Different substrates in cultivation media were modeled by changing the uptake bounds of the corresponding substrates. Different gene knockouts were modeled by setting the lower and upper bounds of the respective enzymatic reactions to zero. We examined substrate uptake rates $v_s$ between 0 and 50 mmol/h/grams of cell dry weight (mmol/h/gDW). The linear-fractional optimization problem in Equation (2) was solved by sampling the allowed values $v_s$, (60 samples in equal distances between 0 and 50), maximizing $v_{bm}$ in each case, and selecting the value for $v_s$ that maximizes $\eta$. The analysis was performed using COBRAToolbox [2] and the Linear Programming problems were solved using glpk (http://www.gnu.org/software/glpk/).

## 3. Properties of growth efficiency

### 3.1 Maximum of growth efficiency

The growth efficiency function $H(v_s)$ obtains its maximum at a finite substrate uptake rate $v_s$. This is illustrated in Figure 1 which shows the predicted growth rate and growth efficiency for a wild-type *E. coli* strain in glucose minimal media assuming varying glucose uptake rates. While the growth rate increases monotonically as the function of the glucose uptake rate, the growth efficiency has a maximum at $v_s^* = 9.2$ mmol/h/gDW. If the uptake rate is greater than $v_s^*$, the cell starts to secrete increasing amounts excess metabolites like acetate.

## 3.2 Sensitivity of growth efficiency to model uncertainties

Metabolic network models are based on well known and validated information on stoichiometric coefficients in biochemical reactions. However, the models also include specific uncertainties. We examined the robustness of the growth efficiency function against four model parameters: (1) the maximal oxygen uptake rate, (2) ATP requirement for growth associated maintenance (GAM), (3) ATP requirement for non-growth associated maintenance (NGAM), and (4) the phosphorus to oxygen (P/O) ratio that reflects the efficiency of ATP synthesis in the electron transfer chain. These parameters have been identified most critical to the behavior of the iAF1260 model [7]. We first examined the form of the growth efficiency function while varying the maximal oxygen uptake rate between 0 and 50 mmol/h/gDW (the original value being 18.5 mmol/h/gDW). Second, GAM and NGAM were varied for +/- 50% of their original values (59.81 and 8.39 mmol/h/gDW, respectively) by constraining the respective reaction rates. Finally, P/O ratios 0.5, 1.0, 1.75, and 2.67 were tested by modifying the stoichiometric coefficients in the electron transfer chain and constraining specific reactions of the electron transfer chain (similarly as described in [7]). Figure 2 shows that all four parameters affect the growth efficiency function. Oxygen uptake rate has the most drastic effect which shifts $v_s^*$, i.e., the glucose uptake rate at which the maximum of growth efficiency is obtained. GAM and NGAM have similar effects of shifting $v_s^*$ but the effect is more moderate. The increased P/O ratio increases the maximum growth efficiency without notable effects to $v_s^*$.

### 3.3 Substrate uptake rate distributions

Because we assume a bacterium to self-regulate the substrate uptake rate to the maximum $\eta$ at a finite $v_s^*$, we are able to set the substrate uptake rate unconstrained. Usually in FBA it is crucial to constrain the substrate uptake rate properly. Otherwise, as shown in the upper panel of Figure 1, the growth rate simply increases monotonically with increasing substrate input rate. In the following analysis, we simulated 10 different cultivation and 1261 genetic conditions to study the distribution of $v_s^*$. Figure 3 shows the results for phenotypes that are predicted to be viable (i.e., the growth rate is greater than 0.1 h$^{-1}$). For them, the substrate uptake rate $v_s^*$ always remains at a finite range. Typical values for $v_s^*$ are from 5 to 20 mmol/h/gDW. The largest $v_s^*$ is obtained under the knockouts of components of ATP synthase, in particular in pyruvate cultivation. The blockage of ATP synthase requires that the needed ATP is synthesized by other mechanisms, such as glycolysis and the citric acid cycle, which requires a large substrate uptake rate.

## 3.4 The relation of growth efficiency and overflow metabolism

In situations where the maximal growth is achieved, a bacterium may not be able to transform all the substrate efficiently to new biomass but an increasing amount of material is directed to waste. This phenomenon of overflow metabolism has been extensively studied as it is detrimental in industrial applications. In the case of *E. coli*, overflow metabolism directs valuable carbon to acetate production instead of biomass generation. This inhibits growth and it may also disturb product synthesis [20].

We studied the relationship between the maximal growth efficiency and overflow metabolism by simulating all single gene knockouts in the iAF1260 model under varying carbon sources. We found that usually the substrate uptake rate at the maximal growth efficiency ($v_s^*$) equals to the substrate uptake rate at the start of overflow metabolism (i.e., the start of acetate production). There are few exceptions to this rule, for example, when the knockout is directed to specific genes of ATP synthase, pyruvate dehydrogenase, or succinate dehydrogenase. Thus, we reason that the maximal growth efficiency is a concept of its own, and it cannot be directly interpreted as the substrate uptake rate threshold above which overflow metabolism starts.

The use of the maximal growth criterion in growth phenotype simulation may easily produce estimates that are sub-optimal in growth efficiency and likely to express overflow metabolism. In order to illustrate the sub-optimality under the maximal growth criterion, we calculated the loss of growth efficiency using the above-mentioned set of 1261 genetic and 10 environmental conditions. In simulations with the maximal growth criterion, the maximal substrate uptake rate was constrained to 10 mmol/h/gDW. The relative loss in growth efficiency was determined as the growth efficiency under the maximal growth criterion divided by the maximal achievable growth efficiency. Figure 4 summarizes the calculated loss ratios. In the figure, the loss of growth efficiency at maximal growth demonstrates that maximal growth wastes input substrate energy.

## **4. Maximal growth efficiency as a cellular objective**

We examined whether the maximal growth efficiency is a plausible cellular objective for *E. coli* cultivated in a small-scale batch process. Therefore, we used the metabolic model iAF1260 to predict the cellular growth rate assuming *E. coli* maximizes the growth efficiency. The growth predictions were compared to two experimental data sets as follows.

First, we predicted the viability for mutant strains carrying single gene deletions. The predictions were produced for 1117 mutant strains cultivated under glucose minimal media [1]. Each mutant strain in the data set has been experimentally determined to be either viable or inviable. For 982 viable mutants, the viability was correctly predicted (i.e. True Positive rate was 97%) and, for 76 inviable mutants, the inviability was correctly predicted (True Negative rate was 72%). The results are identical with the prediction results obtained using the maximal growth criterion.

Second, we predicted the growth rate for 5,096 growth conditions, consisting of 91 single gene knockout strains cultivated under 56 different media conditions. The optical density (OD) of *E. coli* grown in these conditions has been measured in a high-throughput experimental screen using the Biolog platform (http://www.biolog.com), and the data has been set available through the ASAP database [10]. Figure 5 shows the growth predictions for each of the growth conditions versus the corresponding OD value. Under the maximal growth criterion, the Spearman correlation between the predicted growth rates and the experimental OD values was 0.19 while under the maximal growth efficiency criterion the Spearman correlation was 0.24. We also fitted linear models to both data in order to further compare the phenotype prediction performance of the two criteria. Under the maximal growth criterion, the linear model had residual standard error of 0.83 and the adjusted $R^2$ was 0.59. Under the maximal growth efficiency criterion, the linear model had residual standard error of 0.76 and the adjusted $R^2$ was 0.65. The better fit in the case of the maximal growth efficiency was confirmed using Akaike's information criterion (12,537.83 for maximal growth criterion versus 11,639.86 for maximal growth efficiency criterion).

## **5. Conclusions**

The identification of cellular survival strategies and their simulation by realistic objective functions have fundamental importance on phenotype prediction in metabolic analysis. It is probable that there is no single survival strategy that is optimal in all situations but the strategy is likely to depend on growth conditions of a microorganism [8]. Feist and Palsson discuss three qualitatively different environments: nutritionally rich, nutritionally scarce, and elementally limited [8]. Nutritionally rich laboratory-like conditions are probably very rare in the nature and thus, maximal growth is probably an unrealistic objective function in most of the situations. In a study by Schuetz et al. [18] it was found that under nutrient scarcity in continuous cultivations, the best prediction accuracy was achieved using linear maximization of ATP or biomass yields. On the other hand, in unlimited growth on glucose in oxygen or nitrogen respiring batch cultures, the best prediction accuracy was achieved by nonlinear maximization of the ATP yield per flux unit.

In this work we introduced a concept called growth efficiency and characterized its properties. The study was performed using the metabolic model iAF1260 for *Eschericia coli*. As a result we found that the growth efficiency function has its maximum within a finite substrate uptake rate. According to our predictions, the substrate uptake rate at which the maximal growth efficiency is obtained ($v_s^*$) varies typically from 5 to 20 mmol/h/gDW. Our simulations with several different cultivation media and a set of single gene knockouts demonstrated that the optimal rate $v_s^*$ depends on the cultivation and genetic conditions. For example, with sucrose the median uptake rate of $v_s^*$ was 4.5 mmol/h/gDW while with pyruvate the median rate was 17 mmol/h/gDW. We also found that the growth efficiency function is affected by specific parameters that usually remain unsure in metabolic network models. In particular, oxygen uptake and ATP requirement for growth associated maintenance affect $v_s^*$, and increasing P/O ratio in electron transfer chain increases growth efficiency while maintaining the form of the growth efficiency function.

A straight-forward application of growth efficiency is to use it as an optimization criterion (i.e., objective function) for predictions of cellular growth. We explored this possibility and validated our computational predictions using two sets of experimental data. We found that maximal growth efficiency can be considered as a feasible optimization criterion in metabolic modeling. The criterion predicted the given experimental data slightly better than the commonly applied maximal growth rate criterion.

In this study we used data from batch cultivations to validate the feasibility of growth efficiency as an objective function. However, based on the work by Schuetz et al. [18], we hypothesize that the growth efficiency criterion could perform better in situations where cells are under nutrient scarcity, i.e. they are cultivated in nutrient-limiting chemostats. Such chemostat data was not available in this study, and the hypothesis should be validated in a future study.

Considering maximal growth efficiency as a cellular objective suggests that cells can save nutrients in the benefit of other cells or to be used to themselves at a later moment. This raises the question about the mechanisms, e.g. quorum sensing, which bacteria growing in colonies may use to tune their growth rate in each growth situation. As a further point we note that if it proves true that bacterial cells maximize the growth efficiency per unit food or energy uptake, this picks out an optimal rate of energy utilization, hence an optimal displacement from chemical equilibrium for non-equilibrium living cells. We note that we lack the theory of an optimal displacement from equilibrium for living, non-equilibrium, cells. Jacques Monod, in Chance and Necessity, notes that optimally growing bacteria give off little heat [14]. This may be consonant with maximal growth efficiency, so that the maximal amount of energy coming into cells goes into biomass production and minimizes waste heat.

## **Acknowledgments**

This work was supported by the Academy of Finland (Finnish Programme for Centres of Excellence in Research 2006-2011) and the FiDiPro programme of Finnish Funding Agency for Technology and Innovation.

## **References**

[1] T. Baba, T. Ara, M. Hasegawa, Y. Takai, Y. Okumura, M. Baba, K.A. Datsenko, M. Tomita, B.L. Wanner, H. Mori, Construction of *Escherichia coli* K-12 in-frame, single-gene knockout mutants: the Keio collection, Mol. Syst. Biol. 2 (2006) 2006.0008.

285  [2] S.A. Becker, A.M. Feist, M.L. Mo, G. Hannum, B.Ø. Palsson, M.J. Herrgard,
286  Quantitative prediction of cellular metabolism with constraint-based models: The
287  COBRA Toolbox, Nat. Protocols 2 (2007) 727-738.

288

289  [3] L.M. Blank, L. Kuepfer, U. Sauer, Large-scale $^{13}$C-flux analysis reveals mechanistic
290  principles of metabolic network robustness to null mutations in yeast, Genome Biol. 6
291  (2005) R49.

292

293  [4] H.P.J. Bonarius, V. Hatzimanikatis, K.P.H. Meesters, C.D. de Gooijer, G. Schmid, J.
294  Tramper, Metabolic flux analysis of hybridoma cells in different culture media using
295  mass balances, Biotechnol. Bioeng. 50 (1996) 299–318.

296

297  [5] M. Dauner, U. Sauer, Stoichiometric growth model for riboflavin-producing *Bacillus
298  subtilis*, Biotechnol. Bioeng. 76 (2001) 132–143.

299

300  [6] O. Ebenhoh, R. Heinrich, Evolutionary optimization of metabolic pathways.
301  Theoretical reconstruction of the stoichiometry of ATP and NADH producing systems,
302  Bull. Math. Biol. 63 (2001) 21–55.

303

304  [7] A.M. Feist, C.S. Henry, J.L. Reed, M. Krummenacker, A.R. Joyce, P.D. Karp, L.J.
305  Broadbelt, V. Hatzimanikatis, B.Ø. Palsson, A genome-scale metabolic reconstruction for
306  *Escherichia coli* K-12 MG1655 that accounts for 1260 ORFs and thermodynamic
307  information, Mol. Syst. Biol. 3 (2007) 121.

[8] A.M. Feist, B.Ø. Palsson, The biomass objective function, Curr. Opin. Microbiol., 13 (2010) 344-349.

[9] D.A. Fell, J.R. Small, Fat synthesis in adipose tissue. An examination of stoichiometric constraints, Biochem J 238 (1986) 781–786.

[10] J.D. Glasner, P. Liss, G. Plunkett 3rd, A. Darling, T. Prasad, M. Rusch, A. Byrnes, M. Gilson, B. Biehl, F.R. Blattner, N.T. Perna, ASAP, a systematic annotation package for community analysis of genomes, Nucleic Acids Res. 31 (2003) 147-151.

[11] R. Heinrich, F. Montero, E. Klipp, T.G. Waddell, E. Melendez-Hevia, Theoretical approaches to the evolutionary optimization of glycolysis: thermodynamic and kinetic constraints, Eur. J. Biochem. 243 (1997) 191–201.

[12] A.L. Knorr, R. Jain, R. Srivastava, Bayesian-based selection of metabolic objective functions, Bioinformatics 23 (2007) 351–357.

[13] E. Meléndez-Hevia, A. Isidoro, The game of the pentose phosphate cycle, J. Theor. Biol. 117 (1985) 251-263.

[14] J. Monod, Chance and Necessity: An Essay on the Natural Philosophy of Modern Biology, first ed., Knopf, New York, 1971.

331

[15] A.P. Oliveira, J. Nielsen, J. Forster, Modeling *Lactococcus lactis* using a genome-scale flux model,. BMC Microbiol 5 (2005) 39.

[16] N.D. Price, J.L. Reed, B.Ø. Palsson, Genome-scale models of microbial cells: evaluating the consequences of constraints, Nature Rev. Microbiol. 2 (2004) 886-897.

[17] R. Ramakrishna, J.S. Edwards, A. McCulloch, B.Ø. Palsson, Flux-balance analysis of mitochondrial energy metabolism: consequences of systemic stoichiometric constraints, Am. J. Physiol. Regul. Integr. Comp. Physiol. 280 (2001) R695-R704.

[18] R. Schuetz, L. Kuepfer, U. Sauer, Systematic evaluation of objective functions for predicting intracellular fluxes in *Escherichia coli*, Mol. Syst. Biol. 3 (2007) 119.

[19] W.M. van Gulik, J.J. Heijnen, A metabolic network stoichiometry analysis of microbial growth and production formation, Biotech. Bioeng. 48 (1995) 681-698.

[20] K. Valgepea, K. Adamberg, R. Nahku, P.-J. Lahtvee, L. Arike, R. Vilu, Systems biology approach reveals that overflow metabolism of acetate in *Escherichia coli* is triggered by carbon catabolite repression of acetyl-CoA synthetase, BMC Syst. Biol. 4 (2010) 166.

353 [21] A. Varma, B.Ø. Palsson, Stoichiometric flux balance models quantitatively predict

354 growth and metabolic by-product secretion in wild-type *Escherichia coli* W3110, Appl.

355 Environ. Microbiol. 60 (1994) 3724–3731.

356

357

358 **Figures**

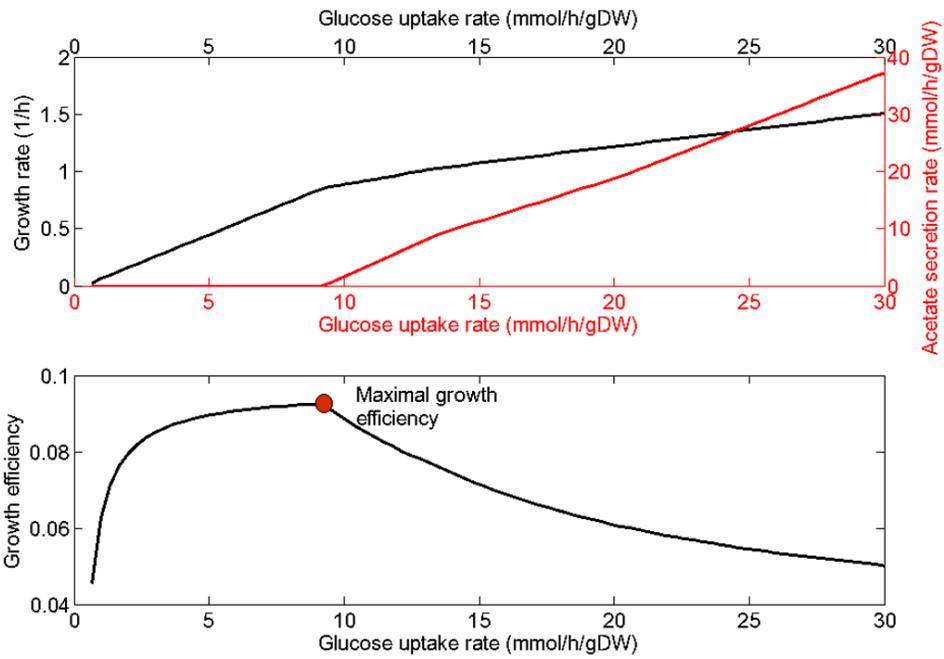

Figure 1. Maximal growth efficiency is obtained at a finite substrate uptake rate. The upper panel depicts the growth rate and the acetate secretion rate as the function of glucose uptake rate. The growth rate is a monotonically increasing function without a maximum. The lower panel shows the growth efficiency as the function of glucose uptake rate, i.e., the growth efficiency function $H(v_s)$.

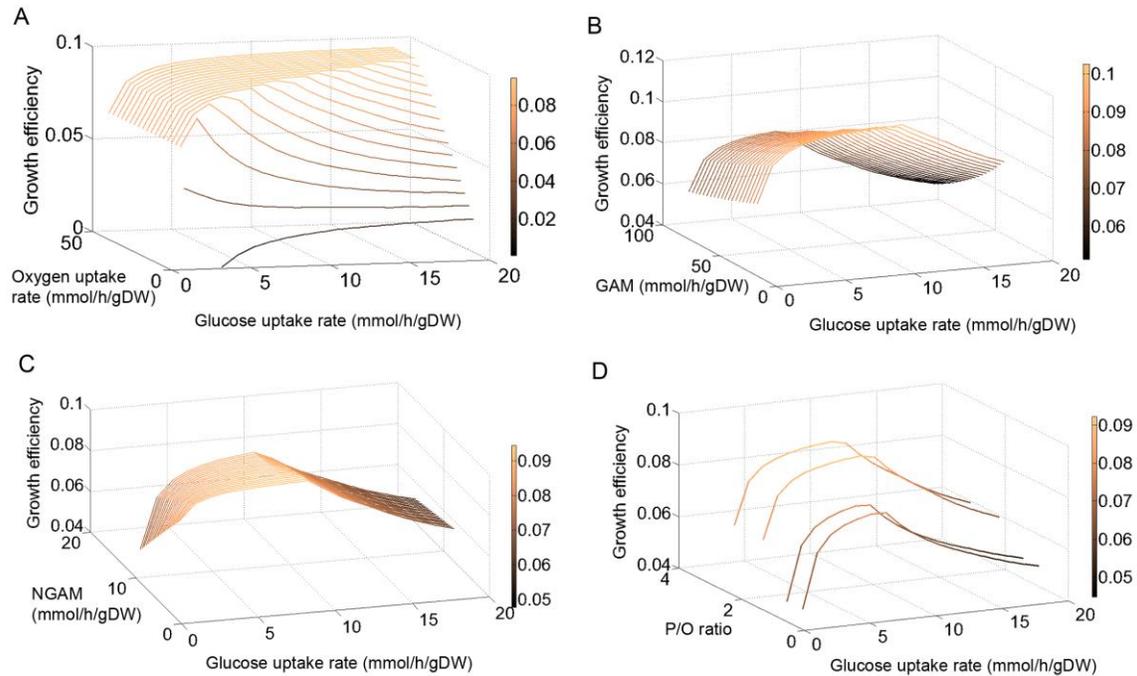

Figure 2. The robustness of the growth efficiency function against four uncertain model parameters. (A) oxygen uptake rate, (B) growth associated maintenance, (C) non-growth associated maintenance, (D) P/O ratio.

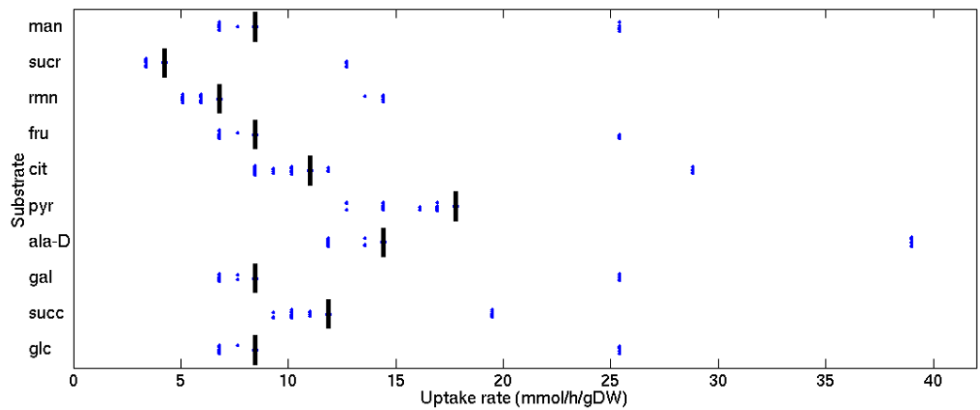

Figure 3. Distributions of substrate uptake rates under the maximal growth efficiency criterion. The uptake rate is calculated using both criteria for 1261 gene knockout strains in 10 carbon sources (glc: glucose, succ: succinate, gal: galactose, ala-D: D-alanine, pyr:

pyruvate, cit: citrate, fru: fructose, rmn: rhamnose, sucr: sucrose, man: mannose). All the distributions are presented as boxplots. The distributions are so concentrated that their data points without outliers appear together as a black bar. Dots represent outliers. Eight data points representing the largest substrate uptake rate (224 mmol/h/gDW for deletions of ATP synthase components in pyruvate cultivation) are not shown.

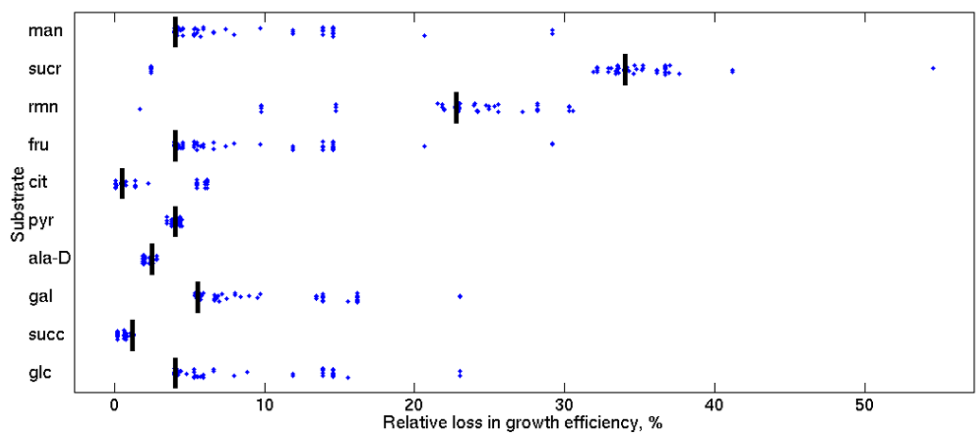

Figure 4. Sub-optimality of growth efficiency under the maximal growth criterion. The relative loss in growth efficiency under the maximal growth rate criterion is depicted for 1261 gene knockout strains in 10 cultivation conditions. The bar represents the median of the 1261 mutants. Almost all the mutants are very close to the median so they cluster under the bar. The distribution is right skewed and the points are at least 2.7 S.D. above or occasionally below the median. The median in almost all culture conditions shows a loss of growth efficiency under the maximal growth rate criterion.

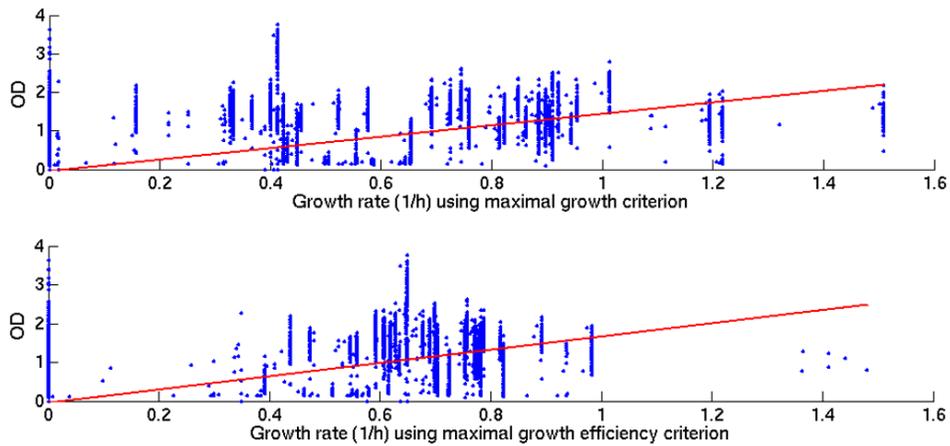

Figure 5. The correspondence of predicted growth rates and experimentally observed optical density. OD values are presented as the function of the predicted growth rates. The upper and the lower panel present the predictions under the maximal growth criterion and the maximal growth efficiency criterion, respectively. A linear model is fitted to both data. In simulations with the maximal growth criterion, the maximal substrate uptake rate was constrained to 10 mmol/h/gDW.